\documentclass{article} 
\usepackage{url,hyperref,graphicx}
\usepackage[square,sort&compress,numbers]{natbib}

\title{Network Robustness Revisited}
\author{Thilo Gross and Laura Barth}
\date{\small Helmholtz Institute for Functional Marine Biodiversity (HIFMB) \\ and  Alfred-Wegener-Institute, Helmholtz Centre for Marine and Polar Research\\ and Carl-von-Ossietzky University, Institute for Chemistry and Biology of the Marine Environment (ICBM)}

\newcommand{\eq}[1]{\begin{equation}#1\end{equation}} 
\newcommand{\eqa}[1]{\begin{eqnarray}#1\end{eqnarray}}

\begin{document}
\maketitle

\begin{abstract}
\noindent{}The robustness of complex networks was one of the first phenomena studied after the inception of network science. However, many contemporary presentations of this theory do not go beyond the original papers. Here we revisit this topic with the aim of providing a deep but didactic introduction. We pay particular attention to some complications in the computation of giant component sizes that are commonly ignored. Following an intuitive procedure, we derive simple formulas that capture the effect of common attack scenarios on arbitrary (configuration model) networks. We hope that this gentle but mathematically-grounded introduction will help new researchers discover this beautiful area of network science. 
\end{abstract}

\section{Introduction}
In 2000 Albert, Jeong, and Barab\'{a}si published a groundbreaking paper on the error and attack tolerance of complex networks \cite{Albert2000error}. At the time of writing this paper has been cited nearly $10^4$ times, and one of the paper's take-home messages, the uncanny stability of scale-free networks, is widely known beyond the academia. Today the study by Albert et al.~is rightfully counted among the founding papers of modern network science. Shortly thereafter, Newman, Strogatz, and Watts published a mathematical theory on the size of connected components in networks with arbitrary degree distribution \cite{newman2001random}. Although some of these results were already known in computer science \cite{molloy1998size}, Newman et al.'s rediscovery popularized them in physics by phrasing them in a convenient and accessible way. Together with other landmark papers published around the same time, these works further accelerated network science which at the time was already rapidly gaining momentum.    

Looking back from the present day, it is clear that several important lines of research directly came from the foundation papers. The mathematics of connectedness after removal some parts of the network, has informed how to prevent catastrophic outages in power grids \cite{albert2004structural,buldyrev2010catastrophic}, fragmentation of communication networks \cite{cohen2001breakdown,doyle2005robust}, cascading species loss in food webs \cite{dunne2002network}, control the outbreak of epidemics \cite{pastor2002immunization,cohen2003efficient,newman2002spread}, detect financial risk \cite{boss2004network,gai2010contagion,haldane2011systemic} and the viral diffusion of misinformation \cite{shao2018spread}. Some important subsequent developments include the extension of the theory to networks with degree correlations \cite{newman2002assortative,vazquez2003assortative},  clustering \cite{berchenko2009emergence,newman2009random}, and block structure 
\cite{priester2014limits}. Moreover structural robustness has been extended to other types of attacks such as cascading failures \cite{motter2002cascade} and bootstrap percolation \cite{watts2002simple,baxter2010bootstrap} and also other classes of systems such multilayer\cite{leicht2009percolation,buldyrev2010catastrophic} and higher order networks \cite{bianconi2018topological}. 

The broad variety of applications makes clear that the theory of network robustness is not the study of an isolated phenomenon, but provides a powerful tool for thinking about network structure. When such new tools are discovered in science they usually go through a phase of tempering where, the underlying mathematics get formulated and subsequently reshaped until a canonical form emerges. For network robustness an important step in this tempering process is the Review by Mark Newman \cite{newman2003structure} which combines known results from graph theory with new approaches to formulate a widely applicable mathematical theory of network robustness. 

Our goal here is not to argue that robustness is the most important topic in network science. There are other topics which were already going strong at the time, and some of them such as network dynamics, and community structure may address a wider range of applications. In fact, to a network scientist it should be apparent that arguing about the relative importance of field is largely meaningless as long as they remain densely interlinked and thus form part of an emergent whole.  

Over the past decades the theory of network robustness has certainly grown into one of the main pillars of modern network science. It is included in several influential reviews and textbooks \cite{callaway2000network,cohen2010complex,newman2018networks,latora2017complex}. However, in current literature, the discussion of robustness does not usually go deeper than Newman's concise presentation. Moreover, there seem to be several very useful corollaries to basic results on robustness, which have not been spelled out in the literature. Finally, while the hallmark robustness of scale-free networks is widely known, the several caveats and flip-sides to this result are known by experts but have received much lesser attention. 

It is our belief that under normal circumstances much more tempering of the theory of network robustness would likely have happened. However, at the time network science was moving extremely fast and a small number of network scientists found themselves suddenly in a position where they could suddenly make a significant impact on a vast range of applications.
In this situation, it was more attractive to go forward to apply and extend the theory rather than to try to rephrase its equations, provide didactic examples, or ponder philosophical issues at its foundations. While all of these things have still happened to some extent, we believe that it is nevertheless valuable to revisit those basic foundations.

The present paper is based on experience gathered while teaching the mathematics of networks robustness over 12 years to different audiences in different departments and on different continents. The paper seeks to provide a retelling of the basic theory that governs the structural robustness of simple networks (configuration model graphs) against different forms of node and link removal. We take the liberty to discuss certain issues at greater length than comparative texts to provide a deep but simple introduction. The presentation is mathematical but, broken into simple steps. We further illustrate the theory by worked examples, including a class of attack scenarios that is exactly solvable with pen and paper. Along the way, we discover some shortcuts and neat equations by which even complicated scenarios can be quickly evaluated. Going beyond mathematics we crystallize the main insights from the calculations into concise take-home messages. We hope that new researchers entering this field will find this introduction of a well-known topic helpful. 

\section{Generating functions}
The exploration of networks builds heavily on the combinatorics of probability distributions. When working with such distributions, we often represent them in the form of sequences 
\eq{
p_k=p_0, p_1, p_2 \ldots 
}
Sequences are intuitive objects, which store information straight forwardly, but they do not come equipped with a lot of powerful machinery. If we want to compute, say, the mean of a distribution, we have to take the elements out of the sequence one-by-one and then process them one-by-one\cite{generatingfunctionology}. By contrast, continuous functions, are mathematical objects that come with a lot of machinery attached; they can be evaluated at different points, inverted, and concatenated. Most importantly, they can be differentiated, enabling us to apply the powerful toolkit of calculus. 

The idea to use functions instead of sequences to store and process distributions lead to the concept of generating functions. An excellent introduction to generating functions can be found in \cite{generatingfunctionology}. In this section, we provide a brief summary of their main properties that are relevant in the context of attacks on networks. 

A sequence can be converted into a function by interpreting it as the sequence of coefficients arising from a Taylor expansion. Applying the Taylor expansion backward turns a sequence $p_k$ into the function 
\eq{
G(x) = \sum_{k=0}^{\infty} p_k x^k.
}
This function is the so-called \emph{generating function} of $p_k$. Note that the variable $x$ does not have any physical meaning, it is merely used as a prop that helps us encode the distribution.

In the following we omit writing the argument of generating functions explicitly if it is just $x$, i.e.~we will refer to the generating function above just as $G$, instead of writing $G(x)$.

\begin{figure}[htb]
    \centering
    \includegraphics[width=0.3\textwidth]{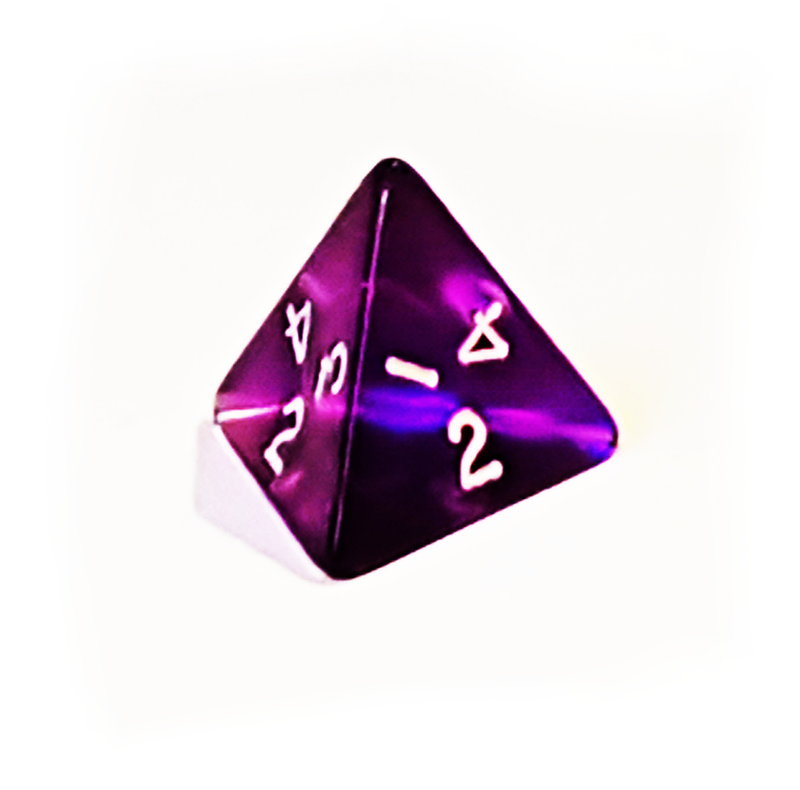}
    \caption{A four-sided die. In contrast to six-sided dice the outcome of a roll is determined by the number of the face on the bottom. The configuration shown in the picture corresponds to an outcome of 2.\label{figDie}}
\end{figure}

For illustration we consider the probability distribution of a (not necessarily fair) four-sided die (see Fig.~\ref{figDie}). We denote the probability of rolling $k$ on a single die roll as $p_k$. Then the generating function for the four-sided die is  
\eq{
G_{\rm 1d4}=p_1 x + p_2 x^2 + p_3 x^3 + p_4 x^4,
}
where we borrowed the notation 1d4 for ``1 four-sided die roll'' that is commonly used in roleplaying games.

\subsection*{Distribution} 
\noindent{}From the generating function we can recover the distribution by a Taylor expansion,
\eq{
p_k = \frac{1}{k!} \left.\left(\frac{\partial}{\partial x}\right)^kG\right|_{x=0}. 
}

\subsection*{Norm} 
\noindent{}In many cases it is unnecessary to recover the sequence as many properties of interest can be computed directly from the generating function. One of these is the norm of $p_k$, which we can compute as  
\eq{
    |p_k| = G(1).
}
For example, for our four-sided dice, we can confirm 
\eq{
G_{\rm 1d4}(1) = p_1 + p_2 + p_3 + p_4.
}

\subsection*{Mean} 
\noindent{}Let's see what happens if we differentiate a generating function. For example,
\eq{
G_{\rm 1d4}' = p_1 + 2 p_2x + 3 p_3 x^2 + 4 p_4 x^3 = \sum_{k=0}^{\infty} k p_k x^k. 
}
The differentiation has put a factor $k$ in front of each of the terms. If we now evaluate this expressing at $x=1$ we arrive at 
\eq{
G_{\rm 1d4}'(1) = p_1 + 2 p_2 + 3p_3 + 4 p_4.
} 
which is the expectation value of the die roll. Also, for any other distribution, we can compute the mean of the distribution as 
\eq{
\langle k \rangle = G'(1).
}

\subsection*{Higher moments}
\noindent{}We can also compute higher moments of the distribution from the generating function in a similar manner. Above, we saw that we can use differentiation to put a prefactor $k$ in front of the terms of the sum in the generating function, however, this also lowered the exponents on the $x$ one count. We can `heal' the exponents after differentiation by multiplying $x$ again, i.e.
\eq{
x \frac{\partial }{\partial x} G = \sum_{k=0}^{\infty} k p_k x^k.
}
Repeating the differentiation and multiplication $n$ times aF prefactor of $k^n$ can be constructed, which allows us to compute
\eq{
\langle k^n \rangle = \sum_{k=0}^{\infty} k^n p_k = \left.\left( x \frac{\partial}{\partial x} \right)^n G \right|_{x=1}.
}

\subsection*{Adding distributions}
\noindent{}Suppose we are interested in the probability distribution of the sum of two rolls of the four-sided die. We could work out the probability for the individual outcomes. For example we can arrive at a result of 4 by rolling a 2 on the first roll and a 2 on the second roll (probability ${p_2}^2$) or a 1 on the first and a 3 on the second ($p_1p_3$) or vice versa ($p_3p_1$) which adds to up to a total probability ${p_2}^2+2p_1p_3$ for a result of 4.

The generating function for the sum of two four-sided die rolls is 
\eq{\begin{array}{r c l}
G_{2d4} &=& {p_1}^2x^2 + 2p_1p_2x^3 + ({p_2}^2+2p_1p_3)x^4 + (2p_1p_4+2p_2p_3)x^5 \\ & & + ({p_3}^2+2p_2p_4)x^6 + 2p_3p_4x^7 + {p_4}^2x^8. \end{array}  
}
Here the first term says that you can get a two by rolling two ones, and so on. 

Looking at the expression for $G_{\rm 2d4}$ it is interesting to note that the combinatorics of the terms is the same that we find in the multiplication of polynomials. This points to a more efficient way for finding $G_{\rm 2d4}$:
\eq{
G_{\rm 2d4} = \left(p_1 x + p_2 x^2 + p_3 x^3 + p_4 x^4\right)^2 = \left(G_{\rm 1d4}\right)^2.
}
So, we can find the generating function for the sum of two die rolls simply as the square of the generating function of one die roll. The same rule holds more generally: Even if we compute the sum of random variables drawn from different distributions, then the generating function for the sum is the product of the generating functions for the parts. 

\subsection*{Adding constants to distributions}
\noindent{}Suppose we want to roll our four-sided die and then add 2 two to the result. We can think of the number 2 as the result of a random process that results in the outcome 2 with 100\% probability. The generating function for such a process is 
\eq{
G_{2} = x^2.
}
We can now use the rule for adding distributions to find the generating function that describes the result of adding two to a four-sided die roll,
\eq{
G_{\rm 1d4+2} = G_2G_{\rm 1d4} = x^2 G_{\rm 1d4}.
}
Generalizing from this result, we can say that when we add $n$ to the outcome of a random process, the generating function that describes the sum is the generating function of the process times $x^n$.  

\subsection*{Adding a random number of random variables (dice of dice)}
\noindent{}Picture a situation in a game where you find a random number of bags, each containing a random number of gold pieces. The player rolls one die to determine the number of bags, and then one die for each bag to determine the gold in that particular bag. The total amount of gold found can then be computed by summing over the values from the individual bags.  
For example, the player might roll a 2 on the first roll, showing that they found 2 bags. Then they roll 1 and 3, finding a single gold piece in the first bag and three in the second for a total of four.

To find the generating function that governs the amount of gold, we could think as follows: With probability $p_1$, we roll a 1 on the first roll, so in this case, we find only one bag. Hence the generating function for the outcome is identical to the generating function of one bag (say, $G_{\rm 1d4}$). With probability $p_2$ ,we roll a 2 on the first roll. Thus, we get two bags and, using the results above, our earnings, in this case, are described by $G_{\rm 2d4}=(G_{1d4})^2$. Putting all four possible scenarios together, we find the generating function for the total amount of gold
\eq{
G_{\rm (1d4)d4} = p_1 G_{\rm 1d4} + p_2 (G_{\rm 1d4})^2 + p_3 (G_{\rm 1d4})^3 + p_4 (G_{\rm 1d4})^4,
}
where the first term corresponds to the scenario where we get one bag, the second corresponds to the scenario where we get two, etc.

\noindent{}Looking at the equation above, we note that it resembles a polynomial of $G_{\rm 1d4}$; we can write it as 
\eq{
G_{\rm (1d4)d4} = G_{\rm 1d4}(G_{\rm 1d4}). 
}
Again, the same rule holds generally: Suppose we have a random process $p$ described by a generating function $P$, and we want to sum over $s$ outcomes of $p$ together, where $s$ is drawn from a distribution with generating function $S$. The generating function for the sum is then 
\eq{
G = S(P).
}

\section{Existence of the giant component}
Large sufficiently-random networks have two distinct phases. In one of these, the network consists of isolated nodes and small components, whereas in the other there are there is a giant component that contains a finite fraction of all nodes, and hence has an infinite size in the limit of large network size \cite{erdos1959graph, erdos1960evolution, erdHos1961strength}. The central question that we review in this paper is how the removal of nodes and links affects the giant component. 

\subsection*{Essential Distributions}
\noindent{}An important starting point for our exploration of giant components is the networks \emph{degree distribution}, i.e.~the probability distribution that a randomly-picked node has $k$ links. We describe this distribution by the sequence $p_k$ and its generating function 
\eq{
G=\sum_{k=0}^{\infty} p_k x^k.
}
the expectation value of the degree distribution is the mean degree 
\eq{
z=\sum_{k=0}^{\infty} k p_k = G'(1). 
}

A second distribution of interest is the \emph{excess degree distribution} $q_k$. If we follow a random link in a random direction, $q_k$ is the probability to arrive at a node that has $k$ links in addition to the one we are traveling on. Finding the excess degree distribution is an example of many calculations in network science that become easier when we think about it in terms of endpoints of links. When we follow a random link (in a random direction) we arrive at a random endpoint. The probability to find $k$ additional links on the node is the same as the probability that a randomly-picked endpoint is on a node of degree $k+1$. Hence we can compute the excess degree distribution as
\eqa{
q_k &=& \frac{\mbox{Number of endpoints on nodes of degree $k+1$}}{\mbox{Number of all endpoints in the network}}  \\   & = &\frac{N(k+1)p_{k+1}}{Nz} = \frac{(k+1)p_{k+1}}{z}.
}
The generating function for this distribution is 
\eq{
Q=\sum_{k=0}^{\infty} \frac{(k+1)p_{k+1}}{z} x^k = \frac{1}{z}\sum_{k=0}^{\infty} kp_k x^{k-1} = \frac{G'}{G'(1)}.
}

The expectation value of the excess degree distribution is the \emph{mean excess degree},
\eq{
q=\sum_{k=0}^{\infty} k q_k = Q'(1) 
=\frac{G''}{G'(1)},
}
i.e.~the expected  number of additional links we find when arriving at a node at the end of a random link.

\subsection*{Existence of the giant component}
\noindent{}In the following, we consider configuration model networks, that is, networks that are formed by randomly connecting nodes of prescribed degree \cite{molloy1995critical,molloy1998size}. 

In such networks, a giant component exists if $q>1$. A mathematical derivation of this result can be found in \cite{newman2001random}. The same result is already derived in principle \cite{molloy1995critical}, but stated in a more complicated and less catchy form, as the concept of excess degree had not been formulated. Here, we skip this derivation of this formula, but, to gain intuition, consider the following argument: if we walk on a network and find on average more than one new link on every node that we visit, we can continue exploring new links until we have seen a finite fraction of the network.  

Despite its intuitive nature, it's good to keep in mind that the $q>1$ condition does not hold in networks that are subject to other organizing principles. Thus it is easy to come up with specific networks that have $q=100$ but no giant component, or a network with $q=0.01$ that has a giant component (see appendix). Although such exceptional networks exist the $q>1$ condition provides a reasonable guide for many real-world applications.         

\subsection*{Size of the giant component}
\noindent{}One of the most subtle and intriguing calculations in network science is determining the size of the giant component. The canonical derivation of this equation starts with a self-consistency statement.
\begin{center}
    \emph{A node is not part of the giant component if none of its neighbors is part of the giant component.}
\end{center}
Note that the statement is phrased in negative form; it is a condition for being outside the giant component, rather than being inside it. One good reason for this formulation is that it makes the equations more concise, as we'll see below.
An unfortunate side effect is that it makes it easier to gloss over a complication that occurs in the next steps.

To arrive at a useful mathematical equation, we need to translate the self-consistency statement into a probabilistic form 
\begin{center}
    \emph{The probability that a randomly picked node is not part of the giant component is the same as the probability that none of its neighbors is part of the giant component.}
\end{center}
We can now assign a symbol to `the probability that a node is not part of the giant component'; say $u$. So, the first half of the statement above says, $u=...$. But what about the second half? It is tempting to jump to the conclusion that for a node of degree $k$, a term of the form $u^k$ will appear. But, let's not go so fast, we first need to deal with some complications. 

\begin{figure}
    \centering
    \includegraphics[width=\textwidth]{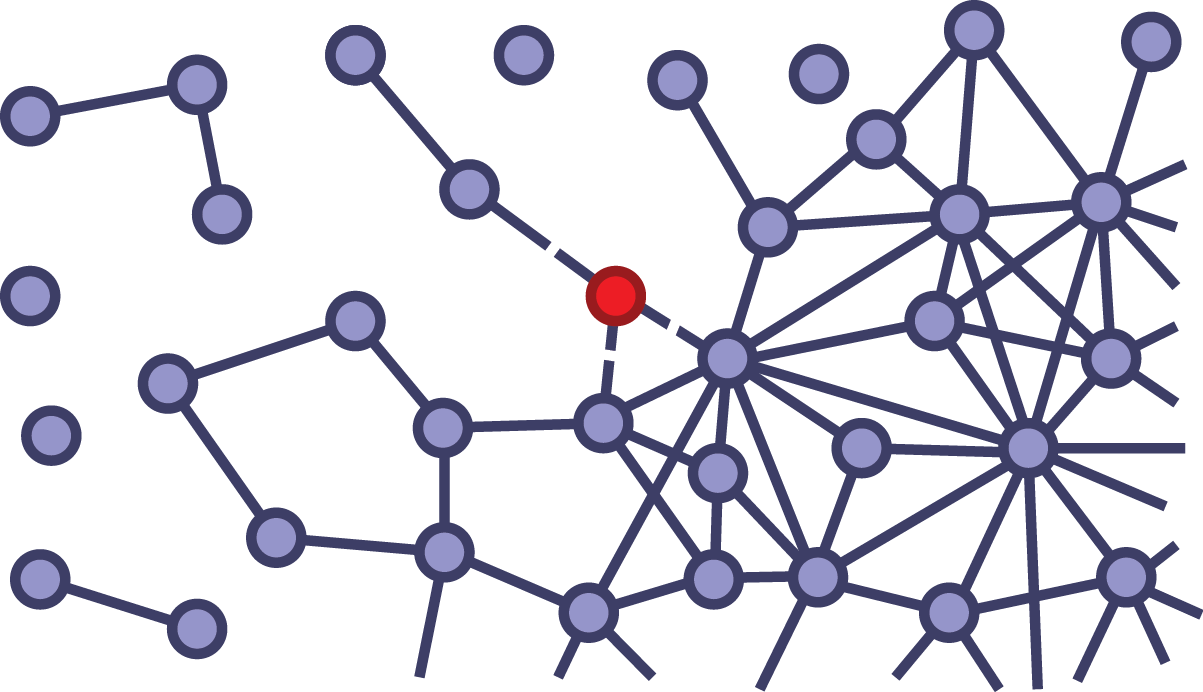}
    \caption{Illustration of the hypothetical cutting of links to find a formula for the giant component size. We pick a random node (red), then cut all of its links. We can say that the probability that the randomly picked node is not in the giant component before the cutting is the same as the probability that none of the node's former neighbors are part of the giant component after the cutting. This statement gives us a self-consistency condition from which the giant component size can be calculated. The cutting of links is essential, as it enables us to treat giant component members of the former neighbors as independent random variables.}
    \label{figCut}
\end{figure}

One problem is that the probabilities in the second half of our statement are not \emph{independent} probabilities. After all, if a node is in the giant component, all of its neighbors must be in the giant component as well. This is bad news because the common mathematical rules for working with probabilities that we often take for granted do not apply. 

For example, if $a$ and $b$ are independent probabilities of events, the probability that both events occur is $ab$, but this isn't necessarily true if the events are interdependent. But if event $b$ must occur if $a$ occurs then, the probability that both occur is just $a$. If we take the interdependence of probabilities into account, our carefully crafted statement above just translates to $u=u$, which would be useless.   

The beauty of mathematical modeling is that by carefully thinking about our definitions, we can often arrive at quantities that work well with mathematics. In the present case, we can use a little twist in the statement to make the probabilities independent:
\begin{center}
    \emph{The probability that a randomly picked node is not part of the giant component is the same as the probability that none of the neighbor's neighbors nodes remain in the giant component after we have removed all of the random node's links.}  
\end{center}
So now we pick a random node, make a list of all of it's neighbors, remove all links from the node and then check whether it's former neighbors are still part of the giant component (Fig.~\ref{figCut}). Because the links to the randomly picked node are broken by the time that we check giant component membership, the probability that the former neighbors are part of the giant component is now independent. 

Having dealt with the issue of interdependence, we could go straight to the solution. However, instead, let us first make an intuitive, but naive attempt. This will lead to a wrong but nevertheless interesting result.

As before, we read the first half of the statement above as $u=\ldots$. To deal with the second half of the statement, we define $v$ as the probability that a given neighbor is not part of the giant component (after the links have been cut). Moreover, let's assume that the degree of our randomly picked node is the mean degree $z$ (for a first attempt, it's worth a try). Under these assumptions, we can translate the statement above to 
\eq{
\label{eqNaiveFirst}
u=v^z.\quad\quad\quad \mbox{\bf [Naive attempt, first half]} 
}
Now we have to ask, what is the probability that $v$ that one of the neighbors is not part of the giant component? If the neighbors were completely random nodes, we could assume $v\approx u$, but we have reached these nodes by the following link. We can now apply the same idea as before: The neighbor is not part of the giant component if none of their neighbors is part of the giant component (after cutting off their links), and hence
\eq{
\label{eqNaiveSecond}
v=v^q.\quad\quad\quad \mbox{\bf [Naive attempt, second half]}
}
Note that the previous Eq.~(\ref{eqNaiveFirst}) links two different variables $u$ and $v$, which appear because a randomly-picked node is statistically different from a randomly picked neighbor. By contrast, the second equation Eq.~(\ref{eqNaiveSecond}) contains two references to $v$ because a random neighbor is statistically similar to a neighbor's neighbor. The second equation is closed, so we can solve it for $v$ and then use $v$ to compute $u$. Using that the proportion of nodes in the giant component is $s=1-u$, we can summarize the solution as follows
\eq{
\label{eqNaiveFinal}
\begin{array}{r c l}
s&=&1-v^z\\ 
v&=&v^q.
\end{array}
\quad\quad\quad \mbox{\bf [Naive attempt, summary]}
}
This was a fund derivation, but unfortunately, the result is now what we wanted from the second equation we can see that the solutions are $v=0$ or $v=1$, which mean $s=0$ or $s=1$, which seems to say, all nodes are in the giant component or none. This can't be right. In addition, there is solution $q=1$, which perhaps hints that something is happening at $q=1$, so perhaps not all is lost?

Thinking about the solutions again, we can see that $s=0$ is a direct consequence of the self-referential nature of our approach: If we just declare every node to be not in the giant component, the result is wrong but self-consistent. Hence it is good to keep in mind that $s=0$ can be a pathological solution that arises from the peculiarities of the approach.   
The situation is worse for the solution $s=1$. This is clearly wrong as our network may well contain some nodes of degree 0 which, certainly can't be in the giant component. Let's understand why we arrive at this erroneous result: In our reasoning, we assumed that every node had the mean degree $z$. By making all nodes the same, we have ended up at a result where all nodes join or leave the giant component together. 

We now understand that the key to a better result is to take the heterogeneity between nodes into account. So instead of assuming that all nodes have the mean degree $z$ or $q$, let's work with the full degree distributions. Our randomly picked node has degree $k$ with probability $p_k$, and, using the same reasoning as above, the neighbors of a node of degree $k$ are not part of the giant component (after link-cutting) with probability $v^k$. 
So that a randomly picked node has degree $k$ \emph{and} is not in the giant component is $p_k v^k$. Similarly, the probability that a randomly picked neighbor has excess degree $k$ and is not in the giant component is $q_k v^k$. Summing over all possibilities for $k$, we find the equations
\eq{
\label{eqImproved}
\begin{array}{r c l}
s&=&1-\sum_{k} p_k v^k\\ 
v&=&\sum_{k} q_k v^k.
\end{array}
\quad\quad\quad \mbox{\bf [Solution]}
}
Examining the form of the solution, we may notice that the generating functions $G$ and $Q$ appear. Hence, we can write the equations for the giant component size as
\eq{
\label{eqGCelegant}
\begin{array}{r c l}
  v&=&Q(v) \\ 
  s&=&1-G(v). 
 \end{array}
 \quad\quad\quad \mbox{\bf [Elegant form of solution]}
}

\subsection*{Degree distribution inside the giant component}
\noindent{}A final ingredient that is sometimes useful is the degree distribution inside the giant component, i.e.~the degree distribution that we would find if all the nodes outside the giant component were removed \cite{molloy1998size,newman2002random}. We already know that the probability that a randomly drawn node has degree $k$ and is not in the giant component is 
\eq{
{p_k}^{\rm out} = p_k v^k.
}
The probability that a node has degree $k$ and is inside the giant component, can be written as 
\eq{
{p_k}^{\rm in} = p_k - {p_k}^{\rm out} = p_k (1-v^k).
}
This probability distribution gives us the probability that a node is inside the giant component and has degree $k$, but what we are interested in is the degree distribution of random nodes picked from the giant component. We can find this by dividing ${p_k}^{\rm in}$ by the probability $s$ that a randomly picked node is in the giant component, which leads to
\eq{
{p_k}^{\rm gc} = \frac{p_k (1-v^k)}{s}.
}
We can use this result to write the generating function for the degree distribution inside the giant component as 
\eq{
G_{\rm gc} = \frac{\sum_{k} p_k(1-v^k)x^k}{s}= \frac{G(x)-G(vx)}{1-G(v)}.
}

\section{Attacks and Damage in Networks}
In the sections above, we established some useful mathematics for estimating the size of the giant component in networks. We are now ready to build a second layer of tools on top of these that capture the effect of different types of attacks and damage in networks. 

\subsection*{Random link removal}
\noindent{}We start by considering an attack that removes links from the network at random. Before the attack, the network is described by the degree distribution $p_k$. Then links are removed at random, such that after the attack, each link survives with probability $c$ (to remember this more easily, we can call this the cir-vival probability). 

We now ask, what is the degree distribution after the attack?
If we were to randomly pick a node from the network after the attack, the probability to pick a node that had $k$ links before the attack is $p_k$. 
Each of these links has a chance $c$ to survive the attack. We can also describe the survival in terms of a probability distribution. A link that was one link before the attack is still one link after the attack, with probability $c$, and it is zero links with probability $1-c$. So the degree of a randomly-picked node, after the attack, is computed as a sum over a random number of random variables. This is exactly the sort of calculation that is covered by the ``dice of dice'' rule from Sec.~2. 

To apply the dice-of-dice rule, we need to describe the attack by a generating function,
\eq{
A=(1-c)x^0+cx^1 = 1+(x-1)c,   
}
which means 1 with probability c and 0 with probability $1-c$. Using the dice-of-dice rule we can then write the degree generating function after the attack as 
\eq{
\label{eqGlinkremoval}
G_{\rm a} = G(A).
}
This equation is a powerful tool, allowing us to derive some results very quickly. For example, we can compute the mean degree after the attack as
\eq{
z_{\rm a}=G_{\rm a}'(1) = G'(A(1))A'(1)=cG'(1)= cz,   
}
where we used the normalization condition $A(1)=1$. This ``norm reduction'' step is a staple of generating function calculations and is one of the reasons why these calculations are often enjoyable. Here, the result shows that removing a proportion of the links reduces the mean degree by the same proportion, regardless of the degree distribution. 

Similarly, we can find the generating function of the excess degree distribution after the attack $Q_{\rm a}$ by substituting the attack function $A$ into $Q$,
\eq{
\label{eqQlinkremoval}
Q_{\rm a}=Q(A).
}
Using generating functions, we can prove this rule in a single line,
\eq{
Q_{\rm a}=\frac{G'_{\rm a}}{G'_{\rm a}(1)}=\frac{G'(A)A'}{G'(A(1))A'(1)} = \frac{G'(A)}{G'(A(1))} = Q(A),
}
where we used $A'(x)=A'(1)$, a property of the attack function. The mean excess degree after the attack is 
\eq{
\label{eqrandomremoval}
q_{\rm a} = Q_{\rm a}'(1) = Q'(A(1))A'(1) = Q'(1)c = cq.
}
This shows that, if we remove a proportion of the links at random, then, also the mean excess degree is reduced by the same proportion. 
We can use Eq.~(\ref{eqrandomremoval}) to calculate the proportion of links that need to be removed from a network to break the giant component. Suppose we have a network with excess degree $q$ before the attack and $q_{\rm a}=c q_{\rm b}$ after the attack. The attack will break the giant component, if $q_{\rm a}<1$, which requires $c<1/q$. Hence the proportion $r$ of links we need to remove from the network to break the giant component by random link removal is 
\eq{
r=1-c=1-\frac{1}{q_{\rm b}} = \frac{q-1}{q}.
}
For example, in the early stages of the CoViD-19 pandemic, one infected person infected on average 3 other people. This number is the mean excess degree of the network in which nodes are infected people and links are contacts that have led to infections. If we had managed to remove 2/3 of the links from the transmission network through hygiene and social distancing, it would have broken the giant component on where the virus was spreading and stopped the pandemic in its tracks. Sadly, these numbers are by now woefully outdated due to the evolution of the delta variant, which is much more transmissible.  

The results we derived so far also permit a first glimpse at the stability of heterogeneous networks. Networks that contain different node degrees can have huge mean excess degrees $q$. Hence we can already see that breaking the giant component in such networks may require the removal of a large proportion of the links. For example, if a network has $q=20$, removal of $r=95\%$ of links is required to break the giant component by random link removal.      

To summarize the results from this section, we can say that the network properties after random removal of a proportion $r=1-c$ of the links are
\eq{N_{\rm a} = N } 
\eq{z_{\rm a} = c z }
\eq{q_{\rm a} = c q }
\eq{ G_{\rm a} = G(A) }
\eq{ Q_{\rm a} = Q(A), }
where $A=cx+(1-c)$.

\begin{figure}
    \centering
    \includegraphics[width=\textwidth]{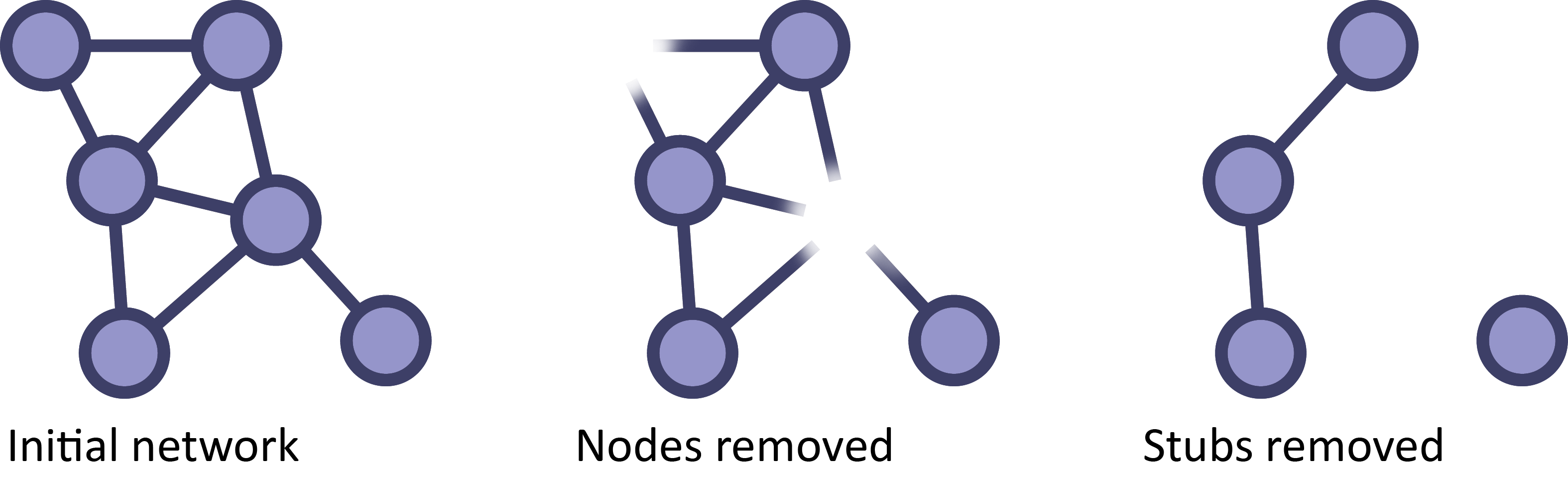}
    \caption{Node removal in a two-step process. Understanding the effect of node removal becomes easier if we picture node removal as a two-step process. Starting from an initial network (left, degrees: 1,2,2,3,4,4) the first step removes the target nodes (here a node of degree 2 and a node of degree 4), but the broken links are kept in the network (center, node degrees 1,2,3,4). In the second step, the broken links are pruned (right, 0,1,1,2). In this example, the mean degree after the first step is $z_{\rm h}=(1+2+3+4)/4=2.5$.}
    \label{figNodeRemoval}
\end{figure}

\subsection*{Random node removal}
\noindent{}Another type of attack on networks is the random removal of nodes. To understand the effect of random node removal, it is useful to imagine it as a two-step process (Fig.~\ref{figNodeRemoval}). In the first step, we remove just the nodes, which leaves behind the broken stubs of links, by which these nodes were connected to the rest of the network. In a second step, we prune these broken links, what may reduce the degrees of the surviving nodes.  

If we remove nodes at random until only a proportion $c$ of the original nodes survives. Already the first step, the removal of the affected nodes, reduces the size of the network. If we had $N$ nodes before the attack, then the number of nodes after the attack is 
\eq{
N_{\rm h} =  c N,
}
where we used the label \textit{h} to indicate that we are now considering the state after the first step, i.e.~halfway through the attack.

Let's also consider what this first step does to nodes of degree $k$. The number of nodes of degree $k$ before the attack is 
\eq{
{n_k} = N p_k.
} 
Since the attack removes nodes at random, a proportion $c$ of the nodes of degree $k$ also survive the first step of the attack, hence
\eq{
{n_k}^{\rm h} = c n_k = c N p_k = N_{\rm h} p_k.
}
We can use this result to compute the degree distribution, after the first step of the attack,
\eq{
{p_k}^{\rm h} = \frac{{n_k}^{\rm h}}{N_{\rm h}} = p_k.
}
This shows that the first step of the attack, the random node removal itself, does not change the degree distribution of the surviving nodes. 

We are not quite done yet, as we still have to clean up the broken links left by the attack. This cleaning up is another example of a calculation that gets easier when we think in terms of endpoints. In the pruning step, a node will lose a given link if the endpoint at the other end of the link was removed in the attack. This means that an attack that removes a certain proportion of all endpoints will remove the same proportion of links from the surviving nodes. Moreover, if we remove a proportion $r$ of the nodes at random, we also remove a proportion $r$ of the endpoints in the system, which implies that in the pruning step we remove a proportion $r$ of the links of the surviving nodes.

Expressed positively, we can say: if a proportion $c$ of the nodes survive, the surviving nodes will retain a proportion $c$ of their links. As the removal of the broken links is essentially random link removal, the same rules as before apply. Thus the mean degree and mean excess degree get reduced by a factor $c$.

In summary, random removal of a proportion $r=1-c$ of the nodes affects the network properties as follows:
\eq{N_{\rm a} = c N } 
\eq{z_{\rm a} = c z }
\eq{q_{\rm a} = c q }
\eq{G_{\rm a} = G(A) }
\eq{Q_{\rm a} = Q(A),}
where $A=cx+(1-c)$.

It is interesting to note that random node removal and random link removal affect the network in very similar ways, which allows us to multiply up the effects of different attacks.

For example, if we vaccinate half the population with a vaccine that is 90\% effective and then also avoid 1/3 of all contacts. We reduce the network mean excess degree, consequently, the remaining vulnerable network is $0.9\cdot 0.5 \cdot 2 / 3 = 0.3$ of its original value. What would certainly, have broken the giant component of the novel Corona Virus wildtype, but insufficient to break the giant component for the delta variant.  

\subsection*{Targeted node removal}
\noindent{}The previous section showed that heterogeneous networks, characterized by high values of $q$, are hard to break by random node removal because we need a proportion of $r=(q-1)/q$ nodes to break the giant component. 

Perhaps we can do better with targeted attacks? The low-dimensional intuition of our daily experience suggests that we can do perhaps much better by attacking naturally existing bottlenecks in the network. A COVID-19 example of this strategy is, for example, trying to stop the virus at national borders; a strategy that has had mixed success. 

When it comes to random networks, our real-world intuition can be misleading: Unless we consider networks of low mean degree, which are fragile in any case, bottlenecks arise only as a result of the low-dimensional embedding of networks, for example, due to geographical constraints \cite{newman2010networks}. The configuration model networks considered here are genuinely high-dimensional structures and thus generally lack strong bottlenecks. While it is possible to fine-tune an attack to split a strongly geographically embedded network, e.g.~the road network, trying to find a similarly optimized attack in a random network is pointless. 

Even in the absence of bottlenecks, we can still maximize the impact of our attack by targeting highly-connected nodes. As in the case of random node removal, we implement the attack in two steps, where the first step removes only the directly affected nodes but leaves the rest of the degree distribution unchanged. Then the leftover stubs will be removed in a second step.

An important decision is how we encode the targeted removal mathematically. Here, we define $r_k$ as the probability that a randomly-picked node from the original network has degree $k$ \emph{and} is subsequently removed in the attack. Most other papers encode targeted attacks in terms of $\rho_k$, the removal risk of a node of degree $k$ which is related to $r_k$ via
\eq{
\rho_k = \frac{r_k}{p_k}.
}
While the definition of $r_k$ seems more complicated,  we will see below that it leads to particularly nice results. 

In actual calculations $r_k$ is quite intuitive as it follows the the same intuition as the degree distribution. Suppose for example the degree distribution of our network was $0.5, 0.25, 0.25$, such that half the nodes where of degree zero. If we wanted to remove $60\%$ of the nodes of degree 2 then $r_k$ would be $0, 0, 0.15$. 

Having familiarized with the $r_k$, let us now consider a degree targeted attack on a general network. As this first step in our calculation, we calculate some properties that quantify the effect of the attack. For this purpose, it is convenient to define the generating function of $r_k$ as
\eq{
  R=\sum_{k=0}^{\infty} r_k x^k.
}

In contrast to the generating functions used so far, the norm of $r_k$ is not 1 but, the proportion of nodes removed in the attack, i.e.
\eq{
  r = 1-c =\sum_{k=0}^{\infty} r_k = R(1), 
}
where $r$ and $c$ are again the removed and surviving proportions of the nodes. 

A second important property is $\tilde{r}$ the proportion of endpoints that are removed directly in the first step of the attack
Recall that $G'(1)=z$ is the mean degree of the nodes in the network. Hence $NG'(1)$ is the number of all endpoints in the network. Analogously, $NR'(1)$ is the number of endpoints that are removed in the first step of the attack. Hence we can compute $\tilde{r}$ as the ratio
\eq{
\label{eqrtilde}
\tilde{r}=\frac{R'(1)N}{G'(1)N}=\frac{R'(1)}{z}.
}
We can now also defined the proportion of surviving endpoints after the first step
\eq{
\tilde{c}=1-\tilde{r}.
}

Let's also have a look at the second derivative of $R$. For the degree generating $G$ the quantity $G''(1)/z$ is the mean excess degree $q$. So by analogy we may call
\eq{
\delta=\frac{R''(1)}{z},
}
the \emph{removed excess degree} by analogy. 

We can now write the degree distribution after the first step (removal of targeted nodes). It is helpful to first write the number of nodes of degree $k$ after the removal 
\eq{
{n_k}^{\rm h}=Np_k - Nr_k = N(p_k - r_k),
}
where we have again used h to denote properties after the first step of the attack. To find the degree distribution after the removal, we have to divide by the remaining number of nodes, which we can write as $Nc$. Hence,
\eq{
{p_k}^{\rm h} = \frac{N({p_k}^{\rm b} - r_k)}{cN} = \frac{p_k - r_k}{c}.
}
The corresponding generating function is 
\eq{
G_{\rm h} = \sum_{k=0}^{\infty} \frac{p_k - r_k}{c} x^k=\frac{G-R}{c}.
}
Using this function, we compute the excess degree generating function after the first step using the relationship $Q=G'/G'(1)$, which implies 
\eq{
Q_{\rm h} = \frac{G_{\rm h}'}{G_{\rm h}'(1)} = \frac{G'-R'}{G'(1)-R'(1)} = \frac{G'-R'}{z \tilde{c}},
}
where we used Eq.~(\ref{eqrtilde}) to replace 
\eq{
G'(1)-R'(1) = z - z \tilde{r} = z(1-\tilde{r}) = z \tilde{c}. 
}

Let's turn to the second step of the attack and remove the remaining stubs of the broken links. We proceed as in the previous case and define the generating function for the probability that a link remains intact  
\eq{
\tilde{A} = \tilde{c}x+1-\tilde{c}, 
}
and then use the dice-of-dice rule to find the degree and excess degree generating function after the attack
\eq{
G_{\rm a}=G_{\rm h}(\tilde{A})=\frac{G(\tilde{A})-R(\tilde{A})}{c} \quad\quad\quad
Q_{\rm a}=Q_{\rm h}(\tilde{A})= \frac{G'(\tilde{A})-R'(\tilde{A})}{z\tilde{c}}.  
\label{eqGFTA}
}

At this point, we already have the generating functions that we need for giant component calculations, but, for completeness, let's also compute the mean degree and mean excess degree after the attack:
\eq{
z_{\rm a}=G_{\rm a}'(1)= \frac{\tilde{A}'(1)}{{c}}\left(G'(\tilde{A}(1))-R'(\tilde{A(1)}\right)=\frac{\tilde{c}}{c} \left(G'(1)-R'(1)\right)=\frac{z\tilde{c}^2}{c},
}
\eq{
q_{\rm a}=Q_{\rm a}'(1)= \frac{\tilde{A}'(1)}{\tilde{c}}\frac{G''(1)-R''(1)}{z} = q-\delta.
}
The second of these equations justifies why we call $\delta$ the removed excess degree. The simplicity of this equation is surprising and probably hints at some deeper insights that may yet be gained. 

In summary, some network properties after a degree-targeted attack described by the attack generating function $R$ are  
\eq{N_{\rm a} = c N }
\eq{z_{\rm a} = z\frac{\tilde{c}^2}{c} }
\eq{q_{\rm a} = q-\delta }
\eq{G_{\rm a} = \frac{G(\tilde{A})-R(\tilde{A})}{c} }
\eq{ Q_{\rm a} = \frac{G'(\tilde{A})-R'(\tilde{A})}{z\tilde{c}},}
where $\tilde{A}=\tilde{c}x+(1-\tilde{c})$, $\tilde{c}=1-R'(1)/z$, $c=1-R(1)$ and $\delta=R''(1)/z$.


\subsection{Viral attacks}
Another interesting class of attacks that we can treat with the same mathematics are ''viral'' attacks that propagate across the same network that they are attacking. Real-world examples include computer viruses and certain infrastructure disruptions such as traffic gridlock and cascading line failure in power grids, but also viral advertising campaigns, etc. Even vaccinations could be turned into viral attacks on an epidemic if we let recipients of the vaccination nominate further recipients. 

When dealing with viral attacks, one potential pitfall is to confuse ourselves by thinking too much about the dynamic nature of the attack. Network science has good methods for dealing with dynamics, but in this paper, we aim to study attacks from a purely structural angle. We will therefore consider the state of the network after the attack has stopped spreading because it can't reach any more nodes.   

If the attack can spread across every link in the network, it will eventually reach every node in the entire component. It is more interesting to consider an attack that can only spread across a certain portion of the links, chosen randomly.
For example, only some roads may have enough traffic flowing along them to allow gridlock to spread. In the following, we call such links that can propagate the attack as \emph{conducting links}.

Now the attack will infect all nodes that it can reach by a path of conducting links. In other words, the attack reaches the entire component in a different version of the network where we count only the conducting links. Because the non-conducting are now ignored, the components in the network of conducting are smaller than the components in our original network, potentially allowing some nodes to escape the attack.  

\begin{figure}[t]
    \includegraphics[width=\textwidth]{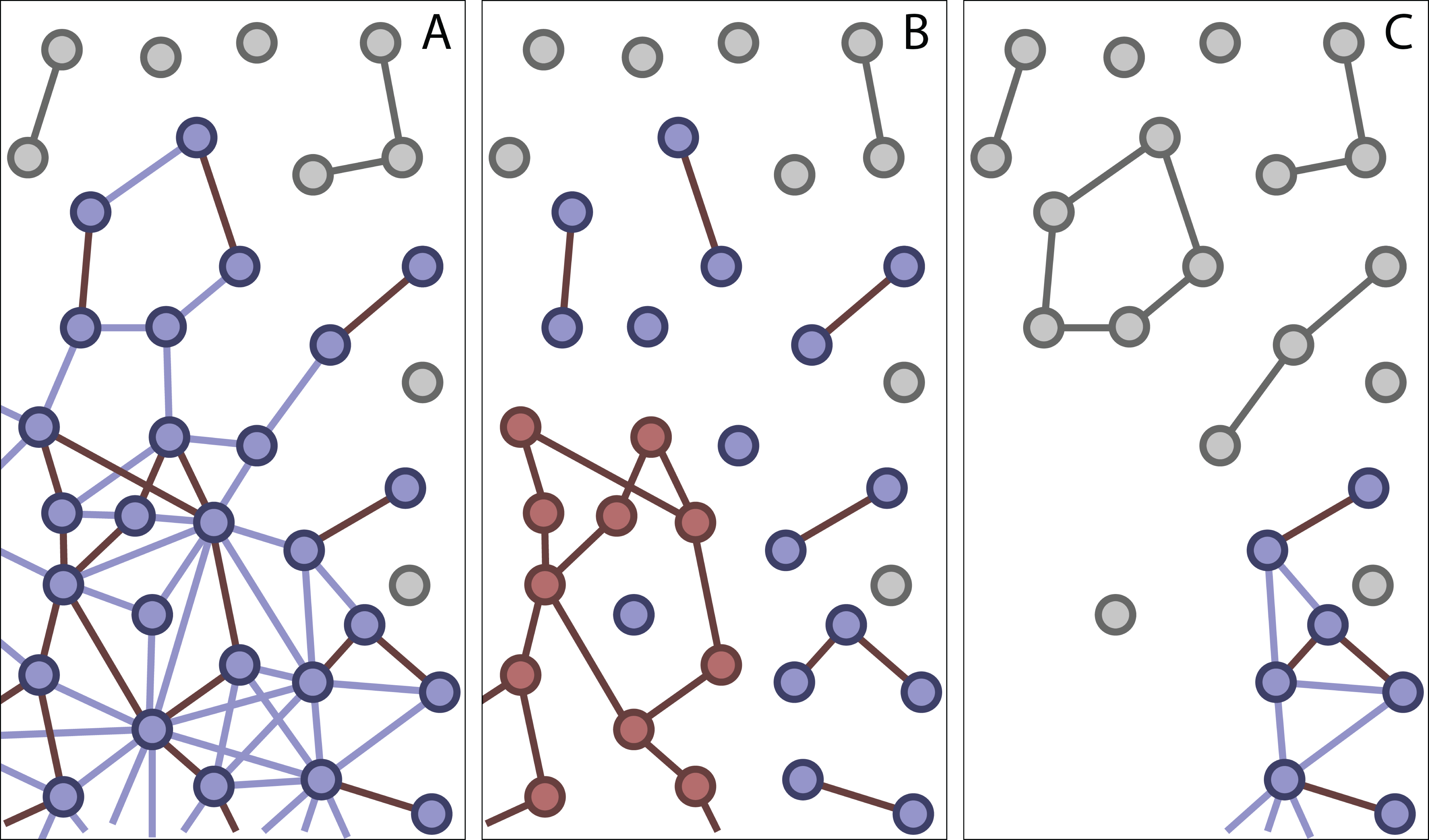}
    \caption{Illustration of a viral attack. Before the attack (A) some proportion of the nodes and links is in the giant component (colored), whereas others are in small components (grey). We consider a situation where only a small fraction of the links conduct the attack (dark red links, only marked in giant component). To assess the impact of a viral attack (B), we remove all non-conducting links and compute the size of the giant conducting component (red nodes). After the attack (C) all nodes in the giant conducting component  and their links have been removed from the original network. The giant component in the remaining network is now smaller as some of its nodes have been destroyed and others have become separated into small components.}
    \label{viral_attack}
\end{figure}

Considering a proportion $w$ of the links as non-conducting is analogous to a link removal attack on the viral attack. Hence if the non-conducting links are distributed randomly, then we can re-purpose our treatment of random link removal to study how many nodes will be affected by the viral attack. For example, we can see immediately that in a network with mean excess degree $q$, there is a giant component in the network of conducting links if $(1-w)q > 1$. Otherwise, a viral attack starting from one node can only spread to a very small number of nodes. 

From now on, we refer to the nodes and links that are part of the giant component in the network of conducting links as the \emph{giant conducting component}.

Furthermore, we can use the results of the random-link-removal attack to compute the number of nodes that are affected by a viral attack. For this purpose, we need to construct a pruning function corresponding to the removal of the non-conducting proportion $w$ of the links,
\eq{
A=(1-w)x+w \label{eqConductingA},
}
which then allows us to compute the giant conducting component size by solving 
\eqa{
v_{\rm c}&=&Q(A(v_{\rm c})) \label{eqconducingB},\\
s_{\rm c}&=&1-G(A(v_{\rm c})). \label{eqGiantConductingComponent}
}
This component size is the proportion of nodes that are removed if an attack starts in the giant conducting component. It is also the probability that a randomly chosen initial spreader will be part of the giant conducting component and hence cause such a large cascade. Otherwise, the initial spreader will be located in a small component of the conducting network and, the attack will only affect a small number of nodes.  

A typical question that arises in the context of viral attacks is if the giant component of the original network can survive a viral attack of a given scale. Thinking about this question becomes much easier if we start in the middle and consider a network in which a certain proportion of links $y$ is \emph{not} in the giant conducting component (Fig~\ref{viral_attack}), either because they are not conducting, or because they are conducting but part of a smaller component. 

We start by constructing the attack generating function $R$ in analogy to our treatment of targeted attacks. If an attack starts in the giant conducting component, it will reach every link except the proportion $y$. Hence a node of degree $k$ will \emph{not} be affected by the attack with probability $y^k$. Conversely, nodes of degree $k$ will be affected by the attack with probability $1-y^k$. Hence the probability that a randomly picked node has degree $k$ and is affected by the attack is 
\eq{
r_k=p_k (1-y^k).
}
Hence the generating function for the node removal is 
\eq{
R=\sum p_k (1-y^k)x^k = G-G(xy).
}
We can now reuse some results from our treatment of degree-targeted attacks. The proportion of nodes affected by the attack is 
\eq{
\label{eqScale}
r=R(1)=1-G(y).
}
The proportion of removed endpoints is 
\eq{
\tilde{r}=\frac{R'(1)}{z}=\frac{G'(1)-yG'(y)}{z}= 1-\frac{yG'(y)}{z},
}
and the reduction in excess degree due to the attack is 
\eq{
\delta = \frac{R''(1)}{z}=q-\frac{y^2G''(y)}{z}. 
}
Hence, after the attack the proportion of remaining nodes is 
\eq{
\label{eqSizeEq}
c=1-r=G(y).
}
The proportion of surviving endpoints is 
\eq{
\tilde{c}=1-\tilde{r}=\frac{yG'(y)}{z}, \label{eqviralendpoints}
}
and the remaining excess degree of the network is
\eq{
\label{eqqviral}
q_{\rm a}=q-\delta = \frac{y^2 G''(y)}{z}.
}
We can now construct the pruning function 
\eq{
\label{eqviralpruning}
\tilde{A}=\tilde{c}x+\tilde{r}.
}
Using Eq.~(\ref{eqGFTA}) we can write the generating functions after the attack
\eqa{
G_{\rm a}&=&\frac{G(\tilde{A})-R(\tilde{A})}{c} = \frac{G(\tilde{A}y)}{G(y)}  \\
Q_{\rm a}&=& \frac{G'(\tilde{A})-R'(\tilde{A})}{z\tilde{c}} = \frac{G'(\tilde{A}y)}{G'(y)},
}
from which we can compute the giant component size in the usual way. 

So far, all of these results are expressed in terms of $y$. Let's explore how $y$ (the proportion of links that are not in the giant conducting component) is related to the more intuitive $w$ (the proportion of non-conducting links). We start by noting that we have two ways to compute the number of nodes removed in the attack on the giant conducting component. We can compute it from our calculation of the giant conducting component size in Eq.~(\ref{eqGiantConductingComponent}). Otherwise we can compute it via Eq.~(\ref{eqScale}) from the attack function $R$. Combining these two equations we get, 
\eq{
G(A(v_{\rm c})) = G(y),
}
since $G$ is a rising function, this implies
\eq{
\label{eqmagicyequation}
y=A(v_{\rm c}),
}
which we can compute from $w$ using Eqs.~(\ref{eqConductingA},\ref{eqconducingB}).

In summary, after a viral attack that can spread across a proportion $1-w$ of the links in the network, will result in a large outbreak with a probability of $1-G(y)$, and if it does, will affect the network as follows: 
\eq{N_{\rm a} = G(y)N }
\eq{z_{\rm a} = \frac{(yG'(y))^2}{zG(y)}}
\eq{q_{\rm a} = \frac{y^2 G''(y)}{z}}  
\eq{G_{\rm a} = \frac{G(\tilde{A}y)}{G(y)}} 
\eq{Q_{\rm a} = \frac{G'(\tilde{A}y)}{G'(y)},}
where $y=A(v_{\rm c})$, $A=(1-w)x+w$, and $v_{\rm c}$ is the solution of $v_{\rm c}=Q(A(v_{\rm c}))$.


\section{Examples and General Results}
The results reviewed in the sections above provide us with a powerful toolkit. We now illustrate this toolkit in a series of examples.  

\subsection{Robustness to random attacks}
\noindent{}Let us start with a three-regular graph, where every node has exactly 3 links. This network is interesting because the property of all networks that suffer random attacks on the three-regular graph can be computed analytically, highlighting it as a great example for teaching. 

Since all nodes in this network have degree three, the degree generating function before the attack is
\eq{
G=x^3,
}
and the corresponding excess degree generating function is
\eq{
Q=\frac{G'}{G'(1)} = x^2,
}
which confirms that, if we follow a random link, we expect to find exactly two additional links at the destination, as it should be. 

Because the mean excess degree is only $q=2$, we can break the giant component already by removing half the links at random, but let's see what happens when we start removing nodes or links at random. Using Eqs.~(\ref{eqGlinkremoval},\ref{eqQlinkremoval}) we know that the generating functions after the attack will be 
\eqa{
G_{\rm a}=G(A)&=&(cx+r)^3, \\ 
Q_{\rm a}=Q(A)&=&(cx+r)^2.
}
To find the giant component size we use Eq.~(28) and
\eq{
   v = Q_{\rm a}(v) = (cv+r)^2. 
}
This is a quadratic polynomial and can be factorized straight forwardly. Alternatively, we can guess that $v=1$ will be a solution and then factor $v-1$ out by polynomial long division. Both ways lead us to 
\eq{
v=\frac{(c-1)^2}{c^2}=\frac{r^2}{c^2},
}
from which we can see in a different way that the $v$ reaches 1 (and consequently the giant component breaks) when we have removed half the links at random (i.e.~$r=c$). Let's focus instead on finding the giant component size when it exists. Again following Eq.~(28) we compute
\eq{\label{homGC}
s=1-G_a(v)=1-(cv+r)^3 = 1-v(cv+r) = 1-v(r+c)r/c = 1-vr/c = 1-r^3/c^3.
}
This result shows that the regular graph is initially quite tough. 
Before we start removing nodes or links, the giant component contains all nodes. For a small attack, the reduction in giant component size initially scales like $r^3$ and hence removing a small proportion of the nodes and/or links has almost no effect on the size of the giant component in the remaining network. But, once a significant proportion of nodes/links have been removed, the impact on the giant component accelerates and quickly leads to its destruction.

Let us compare these results from the regular graph with a network where three-quarters of the nodes have degree 1 and one quarter has degree 9. This network also has a mean degree $z=3$, but its mean excess degree is $q=6$. The generating functions before the attack are
\eqa{
G(x)&=&\frac{3x+x^9}{4}, \\
Q(x)&=&\frac{1+3x^8}{4}.  
}
To find the size of the giant component before the attack, we solve 
\eq{\label{neig}
v=\frac{1}{4}+\frac{3v^8}{4}.
}
While we could solve this equation numerically, an insightful shortcut is to note that the solution must be very close to $v=1/4$. Using this approximate solution, we can then compute the giant component size as follows:
\eq{
s=1-\frac{3v+v^9}{4} = 1-v\left(\frac{9+(4v-1)}{12}\right),
}
where we used $v^8=(4v-1)/3$ to avoid the inaccuracy from raising a numerical approximation to the 9th power. We can see that the factor in the bracket is approximately 0 and hence 
$s=1-3v/4=1-3/16=0.8125$, which is the correct result up to 4 digits of accuracy.

The result shows that, in this heterogeneous network, the giant component contains only about 81\% of the nodes, even before the attack. Conversely, we know that for a network with $q=6$ removal of $5/6\approx 83\% $ of the network is necessary to break the giant component. 

\begin{figure}
    \centering
    \includegraphics[width=0.65\textwidth]{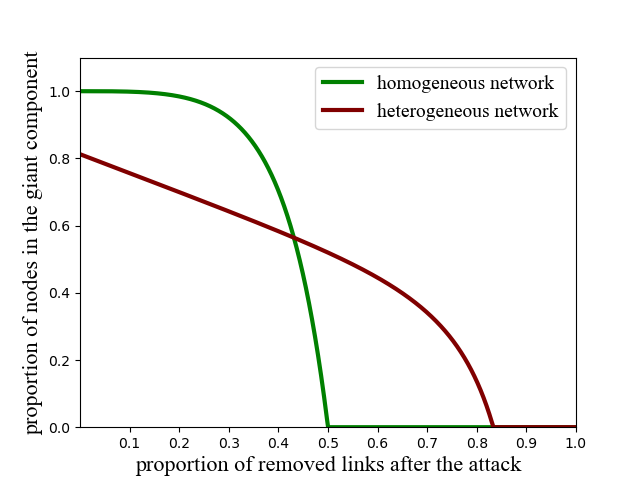}
    \caption{
    Robustness of homogeneous and heterogeneous networks to random damage. Plotted is the proportion of nodes in the giant component versus removed links after the attack for a homogeneous (green, Eq.~\ref{homGC}) and heterogeneous (red, Eq.~\ref{hetGC}). 
    The homogeneous networks resist small attacks better whereas, the heterogeneous network survives a higher proportion of removal. (Random node removal is described by the same curves, in this case, the proportion of the giant component refers to the proportion of remaining nodes).}
    \label{hohe_netwoks}
\end{figure}

To study the effect of the attack in more detail we have to solve
\eq{
v = Q_{\rm a}(v)=Q(A(v))=\frac{1+3(cv+r)^8}{4}, 
}
which we now solve numerically. For teaching (or even a quick implementation on a computer), it is interesting to note that equations of this form can be quickly solved by iteration, i.e.~we interpret the equation as an iteration rule 
\eq{
v_{n+1} = Q_{\rm a}(v)=Q(A(v))=\frac{1+3(cv_n+r)^8}{4}. 
}
Starting from an initial estimate, say $v_0=1/4$, the iteration converges in a few steps due to the high exponent. Once we have obtained the value of $v$ for a given value of $r$, we can compute the corresponding giant component size as
\eq{\label{hetGC}
s=1-G(\tilde{A}(v)),
}
the result is shown in Fig.~\ref{hohe_netwoks}. Although the figure confirms that the giant component persists until 5/6 of the nodes or links have been removed, it also shows that for moderate attacks, the homogeneous topology has a giant component that is larger in absolute terms and also initially less susceptible to attacks. 

This leads us to an important take-home message. We can say, homogeneous networks are like glass: They are very hard when hit lightly but, strong impacts shatter them. Heterogeneous networks are like foam: Parts can be disconnected even without an attack, and it is easy to tear bits off, but it is very tedious to destroy the giant component in its entirety.  

\subsection{Targeted attack on a heterogeneous network}
\noindent{}Let us now consider a targeted attack on the heterogeneous network from the previous section. For a simple start, we explore what happens when we remove half of the nodes of degree 9, i.e.~ we are only removing 1/8 of the total number of nodes in the network. In this case, the generating function for the targeted attack is 
\eq{
R=\frac{1}{8}x^9,
}
and hence we can compute 
\eq{
r=R(1) = \frac{1}{8} \quad\quad\quad s=1-r = \frac{7}{8},
}
\eq{
\tilde{r} = \frac{R'(1)}{z} = \frac{9}{24} \quad\quad\quad \tilde{s}=1-\tilde{r} = \frac{5}{8},
}
\eq{
\delta = \frac{R''(1)}{z} = 3.
}
We can now use the formulas derived above to compute the mean degree and the mean excess degree after the attack
\eq{
z_{\rm a}= z\frac{\tilde{c}^2}{c} = \frac{75}{56} \quad\quad\quad q_{\rm a} = q-\delta = 3.
}
So, in this case, removing 1/8 of the nodes already halves the excess degree. We can also ask what proportion $p$ of the nodes we need to remove if we only target nodes having initially degree 9. We can consider the attack $R=rx^9$. Since we need $q_{a}=1$ to break the giant component and start with $q=6$, 
\eq{
\delta = 5 = \frac{R''(1)}{z} = 24 r.
}
Hence, we can break the giant component by removing $r=5/24$ of the nodes, which is a little bit more than 20\%. 

\begin{figure}
    \centering
    \includegraphics[width=0.65\textwidth]{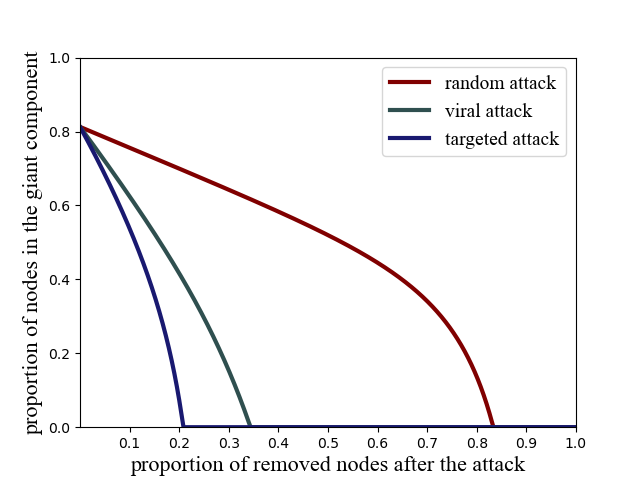}
    \caption{
    Effect of different types of attacks on heterogeneous networks.
    Shown is the giant component size of the heterogeneous example network after a random attack (red, Eq.~\ref{hetGC}), a viral attack (green, Eq.~\ref{eqViralFinal}) and an optimal degree-targeted attack (blue, Eq.~\ref{targ_het}) . Targeting the nodes of highest degree destroys the giant component very quickly. The viral attack is almost as efficient in destroying the network, while requiring much less information on the node degrees.}
    \label{figTargeted}
\end{figure}

We can also compute the size of the giant component after a proportion $r$ of the nodes is removed in an attack that targets only the high degree nodes. Considering again $R=rx^9$, we first compute the proportion of surviving endpoints using Eq.~(\ref{eqrtilde})
\eq{
\tilde{c}=1-\frac{9r}{3} = 1-3r,
}
and the pruning function 
\eq{
\tilde{A} = (1-3r)x+3r.
}
Which allows us to write the self-consistency condition for $v$ as
\eq{
v=Q_{\rm a}(v)=\frac{G'(\tilde{A}(v))-R'(\tilde{A}(v))}{z\tilde{c}} 
=\frac{1 + 3 \left(1 - 4r\right)((1-3r)v+3r)^8}{4(1-3r)},
}
and the giant component size as 
\eq{\label{targ_het}
s=1-G_{\rm a}(v)= 1 -
\frac{G(\tilde{A}(v))-R(\tilde{A}(v))}{c}= 1 -
\frac{3\tilde{A}(v)+(1-4r){\tilde{A}(v)}^9}{4(1-r)}.   
}
This again can be solved by numerical iteration or parametrically. 
A comparison between the effect of the targeted and the random attack on the heterogeneous network is shown in Fig.~\ref{figTargeted}. This illustrates the fragility of heterogeneous networks to targeted attacks \cite{albert2004structural,tiago}. By contrast, the effect of a targeted attack on a homogeneous network is the same as a random attack, as it contains only nodes of the same degree. 


\subsection{A viral attack}
For our final example, we study a viral attack on the heterogeneous example network. For illustration, we consider the case where 80\% of the links are non-conducting, i.e.~w = 0.8.

Following Eq.~(\ref{eqConductingA}), we can prune the none conducting links from the network by the pruning function 
\eq{
A=0.2x+0.8,
}
and hence the generating functions of the conducting network are  
\eqa{
G_{\rm a}&=&G(A) = \frac{3(0.2x+0.8)+(0.2x+0.8)^9}{4} \\ 
Q_{\rm a}&=&Q(A) = \frac{1+3(0.2x+0.8)^8}{4}.
}
We find the giant conducting component by iteratively solving Eq.~(\ref{eqconducingB}):  
\eq{
v=Q_{\rm a}(v),
}
which yields $v\approx 0.735$, and then compute the conducting component size from Eq.~(\ref{eqGiantConductingComponent}),
\eq{
s_c=1-G_{\rm a}(v)\approx 0.136.
}
This tells us that an attack that starts from a randomly-selected node will lead to a large outbreak with a 13.6\% probability, and if it does, it will remove 13.6\% of the nodes. 

To explore the effect that the removal has on the remaining network, we compute $y$ using Eq.~(\ref{eqmagicyequation}),
\eq{
y=A(v)\approx 0.947, 
}
so almost 95\% of links are not in the giant conducting component. 

Now that we know $y$, we can use the Eq.~(\ref{eqviralendpoints}) to compute the proportion of surviving endpoints after the attack,
\eq{
\tilde{c}=\frac{yG'(y)}{z}=\frac{3y+9y^9}{12} \approx 0.697,
}
and the proportion of removed endpoints,
\eq{
\tilde{r}=1-\tilde{c}\approx 0.303.
}
We can now construct our pruning function, $\tilde{A}$, for the viral attack itself (Eq.~\ref{eqviralpruning}) and then compute the giant component size by first solving 
\eq{
v=Q_{\rm a}(v) = Q(\tilde{A}(v)),
}
which yields $v\approx 0.252$. And then computing the remaining giant component size as
\eq{
\label{eqViralFinal}
s=1-G_{\rm a}(v) \approx 0.640.
}
In summary, we have studied an example where only 20\% of the links actually conduct the attack. With so few links, there is only a 13\% chance that it causes a significant outbreak. However, while such an outbreak, if it occurs, removes only 13\% of the nodes, it preferentially hits the nodes of high degree and, as a result, only 64\% of the nodes in the surviving network remain in the giant component. 

Repeating the calculation for different values of $w$ reveals that the viral attack is an intermediate case between random and optimal degree targeted attacks (Fig.~6). In heterogeneous networks, they are almost as damaging as the optimal degree targeted attack while not requiring the attacker to know the complete degree sequence of the network.  


\section{Conclusions and Discussion}
In this paper, we revisited the well-known topic of attacks on networks. We aimed to present this topic in a consistent and didactic way and show that the effect of four types of attacks (random removal of links, random removal of nodes, degree-targeted removal of nodes, viral attacks) can be summarized in compact equations. In many cases these equations, can be solved with pen and paper. 

Our examples illustrate some important and widely-known take-home messages about the robustness of networks. As these are sometimes misconstrued in the wider literature, let us try to restate these messages clearly:
\begin{itemize}
    \item Networks with homogeneous degree distributions are like glass, they are incredibly hard when attacked lightly, but heavier attacks can shatter them easily.
    \item Networks with heterogeneous degree distributions are like foam. Random attacks can quickly detach parts of the giant component. However, shedding the weakest parts enables the giant component to survive significant damage. 
    \item Degree-targeted attacks are relatively pointless against homogeneous networks as the variation in node degrees is low. 
    \item Degree-targeted attacks against heterogeneous networks are devastating and can quickly destroy the giant component. 
    \item Propagating / viral / cascading attacks that spread across the network itself are almost as dangerous as degree targeted attacks as they hit high-degree nodes with high probability.  
\end{itemize}
We emphasize that these are only the most basic insights into configuration-model type networks, and thus strictly hold only in the absence of additional organizing principles such as strong embedding in physical space or the presence of degree correlations and short cycles. Several other papers have extended the theory reviewed here to alleviate these constraints. Notable results include the positive effect of positive degree correlations, which can make the network much more robust against targeted attacks \cite{newman2002assortative,vazquez2003assortative}, and the effect of clustering of short cycles, i.e.~network clustering \cite{berchenko2009emergence,newman2009random}. 

For the class of random and degree-targeted attacks, we showed that the effect of these attacks on the mean and mean excess degree can be captured by very simple equations that can be derived relatively straight-forwardly. Moreover, we pointed out a case (the three-regular-graph) for which the giant component size after all types of attacks can be computed analytically in closed form. For other networks, numerical solutions are needed, but they can be solved by quick numerical iteration on a calculator, rather than requiring full-scale numerics.  

In this paper, we have often referred to the example of vaccination campaigns, and hence a scenario where we want the attack to succeed. However, many of the insights gained can also be applied to make networks more robust against attacks. Many of the conclusions that have been drawn have been discussed abundantly in the literature. Instead of reiterating these, let us point out some issues that have gained comparatively less attention. While it is widely known that the giant component in scale-free networks is highly robust, the results from our examples show that more homogeneous networks are robust in a different way: They resist weaker attacks exceptionally well and are also much less susceptible to targeted and viral attacks.     
It is interesting to reflect on the stability of homogeneous networks against small-scale attacks and damage in a business context. For private businesses, catastrophic events that cause large-scale damage are often not a primary concern, as government actors are expected to intervene in the case of such an event. In comparison, small-damage events typically arrive at a higher rate and will have to be dealt with by the network operator on their own. In this light operating, a very homogeneous network might be in the interest of a business that operates it. However, for governments and the general public optimizing networks in this way, may be detrimental as it leads to low disaster resilience. 

The example illustrates a deeper insight into the nature of network robustness: By adjusting topological properties, we can make networks more resilient against certain types of attacks and damage (cf.~\cite{priester2014limits}). However, unless we increase the overall connectivity, this resilience is usually gained at the cost of increasing vulnerabilities to other attacks. In the real world, where increasing connectivity often comes at a steep price, we can still optimize the robustness by shaping the network such that it can optimally withstand the most likely types of damage. However, care must be taken to make sure we also understand the downsides of such optimization. 

Perhaps a more important conclusion from the present work is that the physics of attacks on networks is a rewarding field of study. The authors greatly enjoyed revisiting the relevant calculations, and the results highlighted here provide a flexible toolkit that, in our opinion, still has large potential to be more widely used in a broad range of fields.  
We hope that readers likewise find this review of the foundations of network robustness helpful and will carry this topic into university curricula and new fields of application. 

\section*{Acknowledgements}
This work was funded by the Ministry for Science and Culture of Lower Saxony (HIFMB project) and the Volkswagen Foundation  (grant number ZN3285)

\section*{Appendix: Beyond the configuration model}
In the main text, we mentioned that the results are only valid for sufficiently random networks, i.e.~those that can well be approximated by the configuration model. In typical applications of this class of networks, we know the degree distribution of the real-world network that we are interested in, but we don't know who is connected to whom. In this case, it is sensible to assume a set of nodes of a given degree, connected by randomly placed links. 

Configuration model results are known to hold well in a wide variety of real-world networks, but they can be wrong in very peculiar network topologies. We mentioned above that there can be networks with mean excess degree $q=100$, which don't have a giant component, and networks with $q=0.01$, which have a giant component. Here we provide examples of these two types of networks. 

Consider a fully connected clique of 101 nodes. Clearly $z=q=100$. If we have multiple copies of such cliques, without links connecting them, the resulting network still has $q=z=100$. Considering such a collection of cliques, we can have a network of arbitrary size that has $q=100$ but does not contain a giant component as each component has size 101. 

To construct a network of $N$ with $q=0.01$ that has a giant component, we can take a proportion of 1/199 of the nodes and connect this proportion in a single cycle. Form the rest of the nodes (198/199) we, form isolated pairs. So the degree distribution is 
\eq{
p_k =  \frac{198}{199} \delta_{k,1} + \frac{1}{199} \delta_{k,2},
}
where $\delta$ is the Kronecker delta. We can confirm that this network has $q=0.01$ as desired. At the same time, the cycle that we have constructed contains $1/199$ of all nodes, hence this component contains a finite proportion of all nodes even in the limit $N\to \infty$ and thus is a giant component.
\end{document}